\documentclass[twocolumn,aps,prl,reprint,amsmath,amssymb]{revtex4-1}
\usepackage{epsfig}
\usepackage{graphicx}
\usepackage{dcolumn}
\usepackage{bm}
\usepackage{multirow}
\usepackage{CJK}
\usepackage{float}

\begin{document}
\begin{CJK*}{GBK}{}
\title{Out-of-time-ordered correlator in non-Hermitian quantum systems}

\author{Liang-Jun Zhai$^1$, Shuai Yin$^2$}
\email{sysuyinshuai@gmail.com}
\affiliation{$^1$The school of mathematics and physics, Jiangsu University of Technology, Changzhou 213001, China}
\affiliation{$^2$Institute for Advanced Study, Tsinghua University, Beijing 100084, China}

\date{\today}

\begin{abstract}
We study the behavior of the out-of-time-ordered correlator (OTOC) in a non-Hermitian quantum Ising system. We show that the OTOC can diagnose not only the ground state exceptional point, which hosts the Yang-Lee edge singularity, but also the \textit{dynamical} exceptional point at the excited state. We find that the evolution of the OTOC in the parity-time symmetric phase can be divided into two stages: in the short-time stage, the OTOC oscillates periodically, and when the parameter is near the ground state exceptional point, this oscillation behavior can be described by both the scaling theory of the $(0+1)$D Yang-Lee edge singularity and the scaling theory of the $(1+1)$D quantum Ising model; while in the long-time stage the OTOC increases exponentially, controlled by the dynamical exceptional point. Possible experimental realizations are then discussed.
\end{abstract}

\maketitle

\textit{Introduction}--- Understanding the nonequilibrium dynamics is one of the most interesting and challenging issues in quantum many-body physics~\cite{Dz,Pol,Rig,Abanin}. Spurred in part by the great progress made in experiments, a lot of new dynamical phenomena have been discovered. These include the eigenstate thermalization hypothesis~\cite{Deu,Sre,Rig1}, many-body localization in disordered systems~\cite{Basko,Oganesyan,Schreiber}, scarred states in the Rydberg atomic systems~\cite{Tur,Tur1,Khe,Lin}, dynamical phase transitions in the time domain~\cite{Heyl,Wei,Mitra1}, and the emergent quantum turbulence after a sudden quench~\cite{Berges2008,Gasenzer2011,Erne2018,Oberthaler2018,Eigen2018}. Accordingly, new quantities are needed to be developed to characterize these new phases. It has been realized that the out-of-time-ordered correlator (OTOC) provides a powerful indicator of the dynamical properties in these dynamical quantum systems~\cite{Larkin,Shenker1,Shenker2}. For instance, in quantum chaotic systems, the OTOC shows an exponential growth with a thermal Lyapunov exponent~\cite{Maldacena}, while in many-body localization systems, the OTOC shows a short-time polynomial increase and a long-time logarithmic increase~\cite{Herq}.

Recently, the OTOC is also used to detect the equilibrium and dynamical phase transitions. It has been shown that the Lyapunov exponent extracted from the OTOC exhibits a maximum around the quantum critical region~\cite{Zhai}. Also, Ref.~\cite{Heyl2019} demonstrates that the OTOC itself can be regarded as an order parameter to probe the equilibrium and dynamical phase transitions~\cite{Duan}. This paradigm has been exploited to determine the universality classes of the Rabi and Dicke models~\cite{Sun}.

On the other hand, the non-Hermitian quantum systems bring new insights in statistical mechanics and condensed matter physics. In statistical mechanics, Yang and Lee paved the way to understand phase transitions by analysing the zeros of the partition function in the complex plane of a symmetry-breaking field~\cite{Yang1952,Lee1952}. Moreover, the Lee-Yang zeros and the Yang-Lee edge singularity (YLES)~\cite{Fisher} have been observed in recent experiments~\cite{Wei,Peng}. Also, the non-Hermitian quantum Hamiltonian provides a prototype to study a class of dissipative phase transitions characterised by the parity-time (PT) symmetry breaking~\cite{Bender,Bender1,Ueda1}. Different from usual quantum phase transitions which occur by tuning a parameter in the Hermitian Hamiltonian, dissipative phase transitions are induced by changing the strength of the dissipation. In condensed matter physics, considerable attentions have been attracted by the topological phases in non-Hermitian systems~\cite{Harari,Fuliang,Fuliang2,Wang,Ueda2}. Moreover, recently, the dynamical phenomena in non-Hermitian quantum systems are also studied~\cite{Yin1,Yin2,Moessner,Ueda3,Xue1,Duj}. Given the important roles played by the OTOC in various Hermitian systems, studies on the OTOC in non-Hermitian systems are urgently called for.

In this paper, we study the behavior of the OTOC in the non-Hermitian quantum Ising model. In this model, the PT-symmetry breaking can happen in some excited state with a critical imaginary field~\cite{Lee1952,Yang1952,Fisher}. We call this point as the \textit{dynamical} exceptional point (DEP). We find that the evolution of the OTOC, which is defined in the PT-symmetric ground state, is strongly affected by the DEP. For the parameter smaller than the DEP, the OTOC shows oscillation behavior; while for the parameter larger than the DEP, the evolution of the OTOC can be seperated into two stages: (i) In the short-time stage, the OTOC oscillates periodically. Moreover, we find that when the parameter is near the ground-state exceptional point (GEP), the OTOC can be described by both the $(0+1)$D YLES scaling theory and the $(1+1)$D Ising scaling theory. From the viewpoint of the $(0+1)$D YLES scaling theory, we find that the OTOC amplitude $A$ changes with the distance to the GEP, $g$, as $A\propto g^{-2}$; while from the viewpoint of the $(1+1)$D Ising scaling theory, we find that $A$ changes with the lattice size $L$ as $A\propto L^{-0.5137}$. (ii) In the long-time stage, the OTOC oscillates with an exponentially increasing amplitude. Accordingly, the OTOC can be employed to detect both the GEP and the DEP.

\begin{figure}[tbp]
\includegraphics[angle=0,scale=0.38]{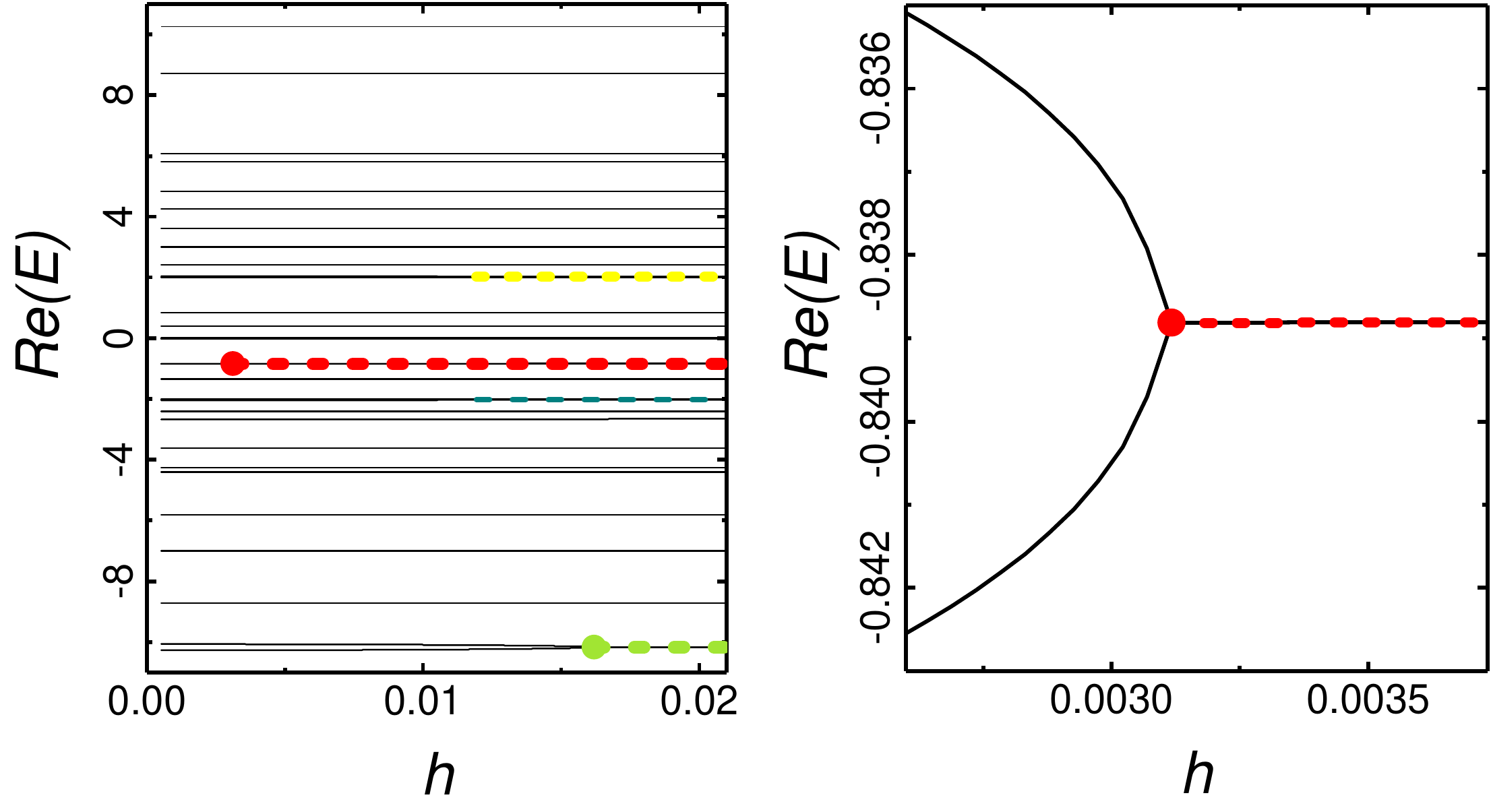}
  \caption{Real-part of the spectra of the non-Hermitian Ising model with $L=8$ and $\lambda=1.0025$. Left: Full spectra in the zero momentum sector. The dashed part indicates the complex spectra. With the increase, the lowest two spectra change from real values to complex values at the GEP (Green dot), reflecting a PT-symmetry breaking phase transition and its universality belongs to the YLES. In addition, in some excited states, complex spectra appear at some DEPs. The spectra with the largest absolute value of imaginary part (Red dashed curve) dominates the long-time dynamics.  Right: Detailed view of the dominating DEP (Red dot). The GEP is at $h_g=0.0162$, and the dominating DEP is at $h_{d}=0.0031$.}
\label{spectra}
\end{figure}

\textit{Hamiltonian and OTOC}--- The Hamiltonian of the system reads
\begin{equation}
    \mathcal{H} = -\sum_i^{L-1} \sigma_i^z \sigma_{i+1}^z - \lambda \sum_i^L \sigma_i^x - i h \sum_i^L \sigma_i^z,
    \label{model}
\end{equation}
in which $\sigma^x$ and $\sigma^z$ are the Pauli matrices in the $x$ and $z$ directions, respectively. The critical point of the ordinary ferromagnetic-paramagnetic phase transition is $\lambda_c=1$ and $h=0$~\cite{Sachdev}, while the YLES occurs at the GEP, denoted as $h_g$, which is the watershed of the PT-symmetric phase with real low-energy spectra and the PT-symmtry breaking phase with complex low-energy spectra~\cite{Bender}. Besides the GEP, one finds that in the excited states, there are some DEPs, denoted as $h_d$, at which the energy spectra change singularly. Periodic boundary condition will be imposed. By using the translational symmetry, we focus on the zero momentum ($k=0$) space in the following. Figure~\ref{spectra} shows the spectra of model~(\ref{model}) for $k=0$. From Fig.~\ref{spectra}, one finds that the parameter at the DEP is smaller than that at the GEP.

The OTOC used in the main text is defined as~\cite{Larkin}
\begin{equation}
     \mathcal{F}(t)= \textrm{Re} \langle G^*|S(t)SS(t)S|G\rangle,
    \label{OTOC}
\end{equation}
in which $S\equiv \sum_i^{L} \sigma_i^z/L$ with $L$ being the lattice size, and $S(t)\equiv \textrm{exp}(i\mathcal{H}t)S\textrm{exp}(-i\mathcal{H}t)$, $|G\rangle$ is the ground state in the PT-symmetric phase. Note that the left vector in Eq.~(\ref{OTOC}) is the transposition of $|G\rangle$. In the supplemental material~\cite{supp}, we also study the OTOC for other initial states.

\begin{figure}[tbp]
\includegraphics[angle=0,scale=0.31]{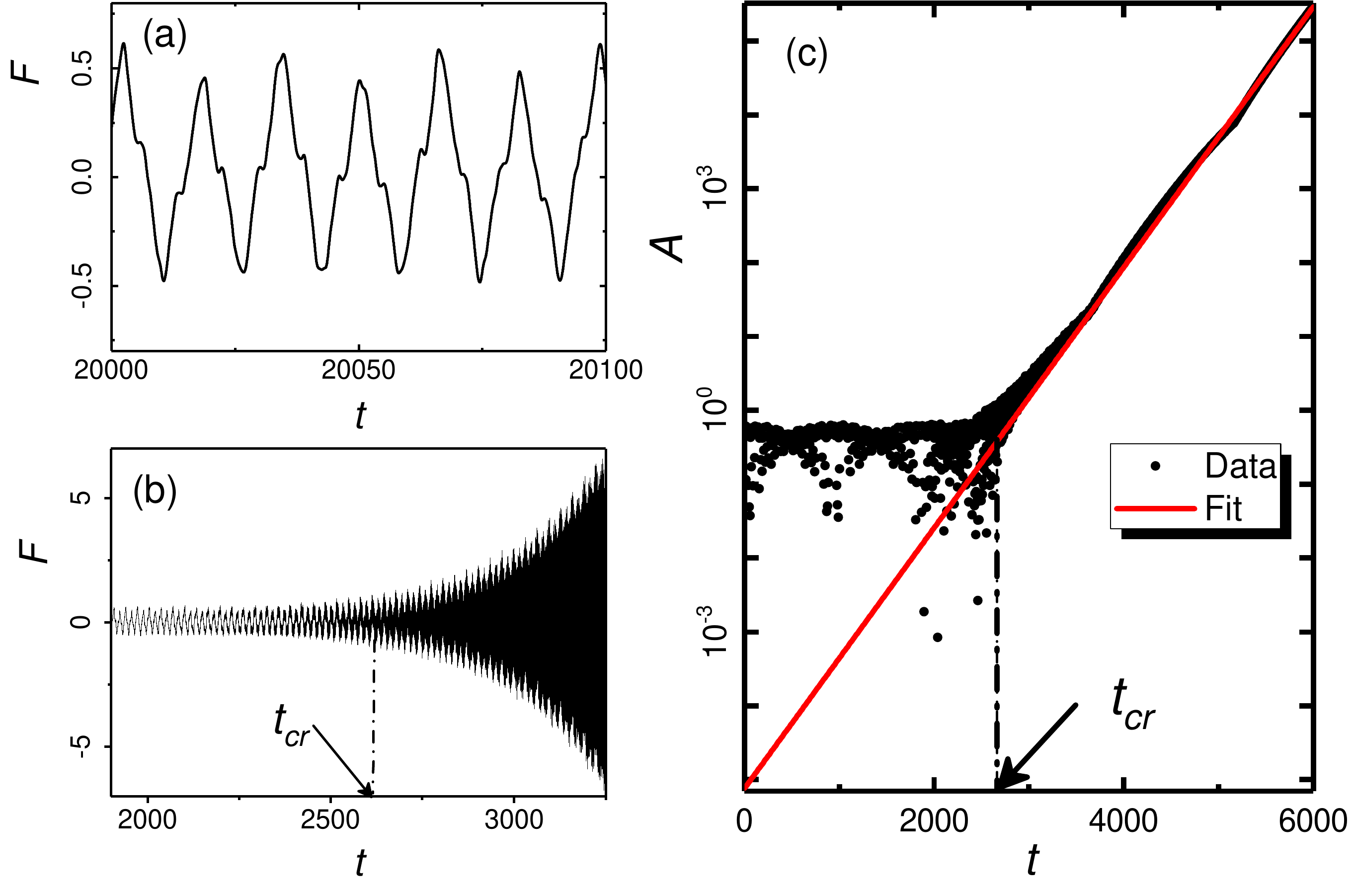}
  \caption{Evolution of the OTOC for $L=8$, $\lambda=1.0025$ and $h_d=0.0031$. (a) For $h<h_d$ ($h=0.00305$), the OTOC oscillates with a period inversely proportional to the lowest energy gap. (b,c) For $h>h_d$ ($h=0.00315$), the evolution of OTOC can be separated into two stages. In the short-time stage, $t<t_{cr}$, the OTOC also oscillates and its period is also inversely proportional to the lowest gap; while in the long-time stage, $t>t_{cr}$, the amplitude $A$ increases exponentially as shown in (c), in which logarithmic coordinate are chosen for vertical axis, and a straight line fit therein indicates a perfect exponential increasement in the long-time scale. $t_{cr}$ is the crossover time as marked.}
\label{general}
\end{figure}

The general feature of the OTOC for model~(\ref{model}) is shown in Fig.~\ref{general}. For $h<h_d$, the OTOC oscillates with a period inversely proportional to the lowest energy gap. This demonstrates that for $h<h_d$, the OTOC is determined by the low-energy physics. Two reasons are listed as follows: (i) $S$ is the summation of the local operator, so the OTOC lives in a space spanned by low-energy states, while the high-energy contributions can be found as the high-frequency fluctuation attached on the period envelope curve; (ii) The evolution operator $\textrm{exp}(-i\mathcal{H}t)$ is similar to a unitary operator since all the spectra are real. In contrast, for $h>h_d$, the evolution can be separated to two stages. In the short-time stage, the OTOC also oscillates periodically, similar to the above case; while in the long-time stage, the OTOC amplitude, $A\equiv |\mathcal{F}|$, increase exponentially. This is because $S|G\rangle$ has non-zero distribution on the excited states, whose energies have imaginary-parts. In the short-time stage, ${\rm exp}({\rm Im}(E)t)$ is small, the OTOC is still controlled by the low-energy physics; while in the long-time stage, $\textrm{exp}(\textrm{Im}(E)t)$ is very large, and the OTOC is controlled by the high-energy physics with largest imaginary-part of the spectra. In the following, we will study in detail the universal dynamic of the OTOC in different stages.

\textit{Short-time dynamics near the GEP}--- The discussion above shows that the low-energy dynamics determines the evolution of the OTOC when $t<t_{cr}$. It has been shown that the low-energy dynamics near the GEP can be described by the scaling theories in both the $(0+1)$D and $(1+1)$D, due to the appearance of the overlapping critical region. This gives rise to a hybrid dynamic scaling theory~\cite{Yin1,Yin2}. Here, we study the scaling behavior of the OTOC near the GEP.

\begin{figure}[tbp]
\includegraphics[angle=0,scale=0.34]{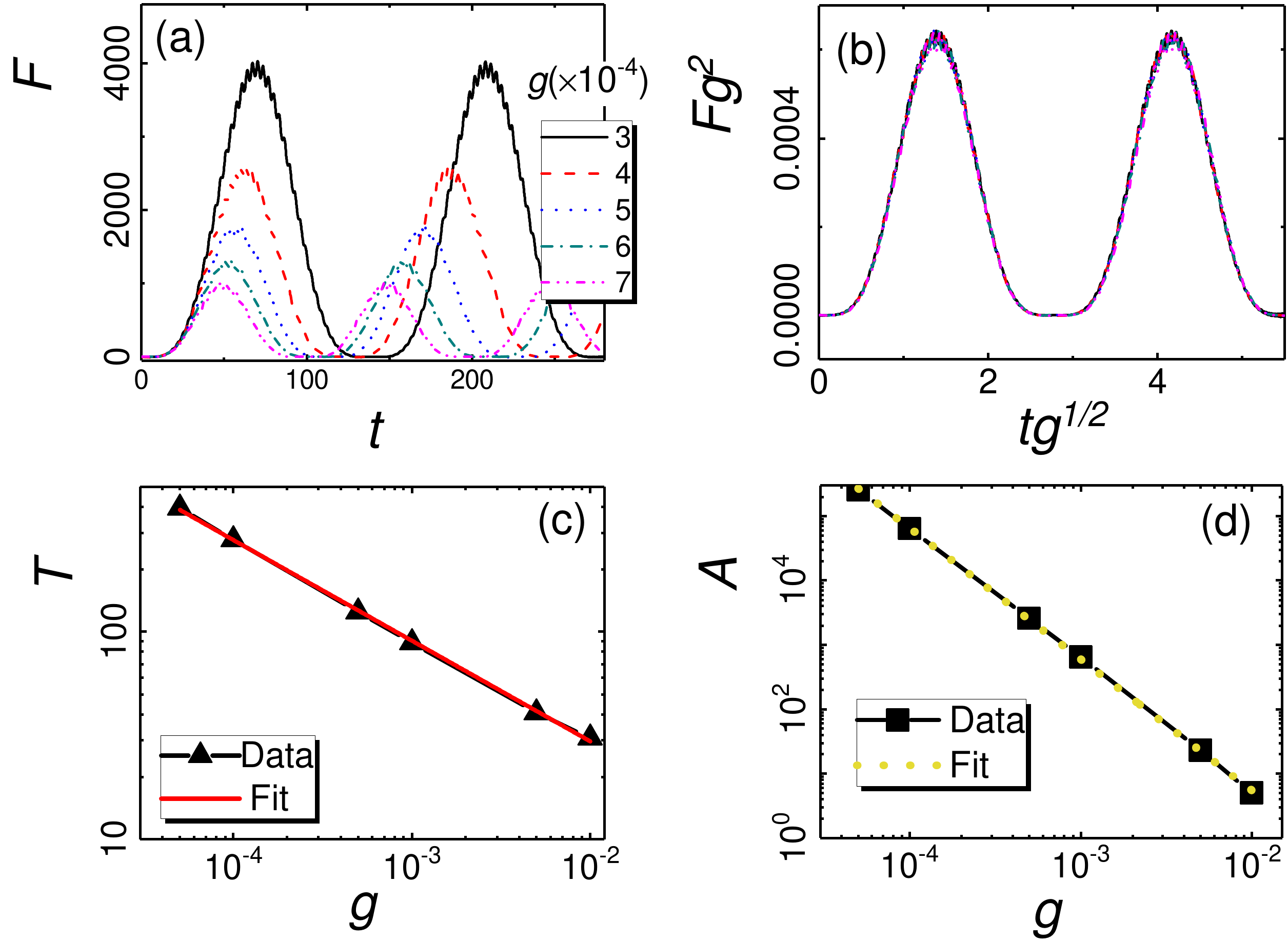}
  \caption{(a) The short-time dynamics of the OTOC for various $g$ with fixed $L=6$, and $\lambda=1.00333$. The corrsponding $h_g$ is at $h_g=0.02785$. (c) The curve for the period $T$ versus $g$. The straight line in the double logarithmic coordinates demonstrates that the relation between them satisfies a power law. The fitting result shows that the exponent is $-0.484$. These results confirm Eq.~(\ref{peri}). (d) The curve for the amplitude $A$ ($A$ is chosen as the value of the first peak) versus $g$. The straight line in the double logarithmic coordinates demonstrates that the relation between them also satisfies a power law. The fitting result shows that the exponent is $-2.039$. These results confirm Eq.~(\ref{ampli}). (b) The rescaled curves collapse onto each other, confirming Eq.~(\ref{zeroscal}).}
\label{zerod}
\end{figure}

We first show that the OTOC can be described by the $(0+1)$D scaling theory of the YLES for the fixed lattice size. We consider the simplest case with $L=1$. In this case, the OTOC can be solved analytically as
\begin{eqnarray}
\begin{split}
\mathcal{F}(t)=&\frac{16}{(\Delta E)^4}\\&\textrm{Re}[h^4+(2-3{\rm e}^{-i\Delta E t}-{\rm e}^{i\Delta E t})h^2+{\rm e}^{-2 i\Delta E t}], \label{twolev}
\end{split}
\end{eqnarray}
in which $\lambda$ has been chosen to be $\lambda=1$, and thus $h_g=1$, and the energy gap $\Delta E=2\sqrt{1-h^2}$, which satisfies with $\Delta E\propto g^{-\nu_0 z_0/\beta_0\delta_0}$, in which $g$ is defined as $g\equiv h_g-h$, and $\nu_0=-1$, $\beta_0=1$, $\delta_0=-2$, and $z_0=1$ are all critical exponents of the $(0+1)$D YLES. Near the GEP, by substituting $h\simeq 1$ into Eq.~(\ref{twolev}), one finds that the OTOC shows oscillation behavior. Its amplitude $A$ diverges as
\begin{equation}
A\propto g^{-2},
\label{ampli}
\end{equation}
while its period $T$ is scaled according to
\begin{equation}
T\propto g^{-1/2}.
\label{peri}
\end{equation}
The latter is consistent with the $(0+1)$D YLES scaling analysis, which gives $T\sim g^{\nu_0z_0/\beta_0\delta_0}$, while the former is a new scaling relation for the OTOC. Because the scaling property is universal near the GEP, it is expected that these two relations are still applicable for the OTOC dynamics in finite-size lattice systems. Furthermore, from Eqs.~(\ref{ampli}) and (\ref{peri}), one finds that the short-time dynamics of the OTOC satisfies,
\begin{equation}
\mathcal{F}(t)=g^{s_0}f_0(tg^{-\nu_0z_0/\beta_0\delta_0}),
\label{zeroscal}
\end{equation}
in which $s_0$ is the $(0+1)$D critical exponent of the OTOC, and  $f_0$ is a scaling function. From Eq.~(\ref{ampli}), one finds that $s_0=-2$. We numerically examine these two relations for model~(\ref{model}). Since $h_d<h_g$ for model~(\ref{model}), we focus on the scaling behavior of the OTOC in the short-time stage. Figure~\ref{zerod} confirms the above two scaling relations. Similar scaling behavior has been recently explored in the Hermitian critical systems~\cite{weibobo}.

We then show that the short-time dynamics of the OTOC can also be described by the $(1+1)$D Ising scaling theory for large enough lattice sizes, which guarantee the usual finite-size scaling is applicable. By taking into account all of the relevant scaling variables, we obtain a scaling function for the OTOC,
\begin{equation}
\mathcal{F}(t)= L^{-s_1/\nu_1} f_1[(\lambda-\lambda_c)L^{1/\nu_1},g_L L^{\beta_1\delta_1/\nu_1},tL^{-z_1}],
\label{isingu}
\end{equation}
in which $g_L$ is $g$ for the corresponding $L$, $\nu_1=1$, $\beta_1=1/8$, $\delta_1=15$, and $z_1=1$ are the critical exponents in the $(1+1)$D Ising universality class, and $s_1$ is a new critical exponent for OTOC in the $(1+1)$D Ising universality class and it is defined as $\mathcal{F}\propto g^{s_1}$.

\begin{figure}[tbp]
\includegraphics[angle=0,scale=0.34]{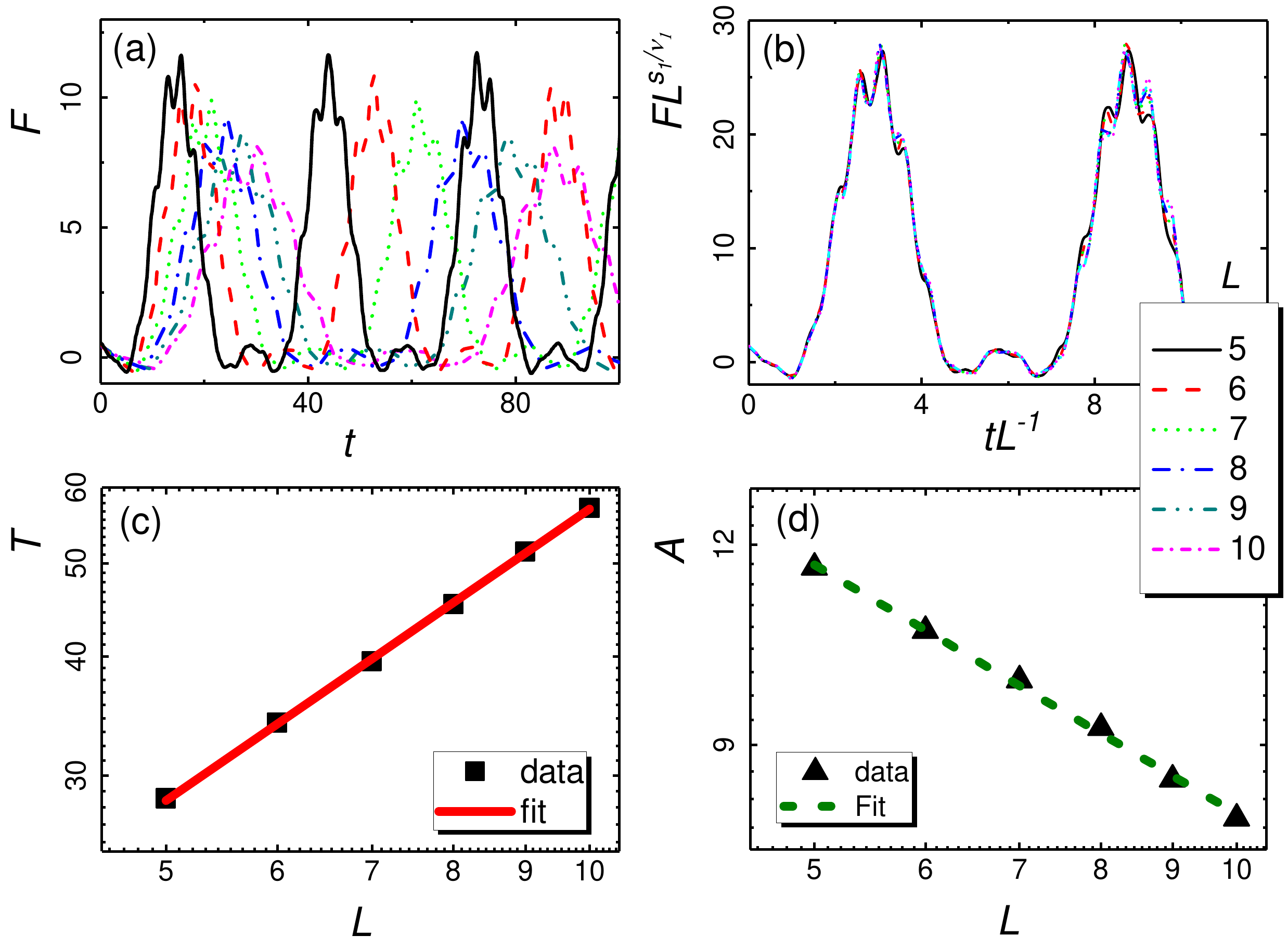}
  \caption{(a) The short-time dynamics of the OTOC for various $L$ with fixed $(\lambda-\lambda_c)L^{1/\nu_1}=0.02$ and $g_L L^{\beta_1\delta_1/\nu_1}=0.20444$. (c) The curve for the amplitude $T$ versus $L$. The straight line in the double logarithmic coordinates demonstrates that the relation between them also satisfies a power law. The fitting result shows that the exponent is $1.013$. These results confirm the scaling variable $tL^{-1}$. (d) The curve for the period $A$ versus $L$. The straight line in the double logarithmic coordinates demonstrates that the relation between them satisfies a power law. The fitting result shows that the exponent is $s_1=0.5137$. (b) The rescaled curves collapse onto each other, confirming Eq.~(\ref{isingu}).}
\label{oned}
\end{figure}

Here, we determine the critical exponent $s_1$ and examine Eq.~(\ref{isingu}). For fixed $(\lambda-\lambda_c)L^{1/\nu_1}$ and $g_L L^{\beta_1\delta_1/\nu_1}$, $\mathcal{F}(t)= L^s f_1^*(tL^{-z_1})$. Since $\mathcal{F}(t)$ oscillates, the dependence of its period $T$ on the lattice size $L$ confirms the scaling variable $tL^{-z_1}$ and the scaling function $f_1$ in Eq.~(\ref{isingu}); and the dependence of its amplitude $A$ on $L$ gives the exponent $s_1$. Accordingly, from Fig.~\ref{oned}(d), we find that $s_1\simeq-0.5137$. Then, by substituting this result into Eq.~(\ref{isingu}), we find that the rescaled curves collapse onto each other, as shown in Fig.~\ref{oned}(b). These results confirm Eq.~(\ref{isingu}).

\begin{figure}[tbp]
\includegraphics[angle=0,scale=0.35]{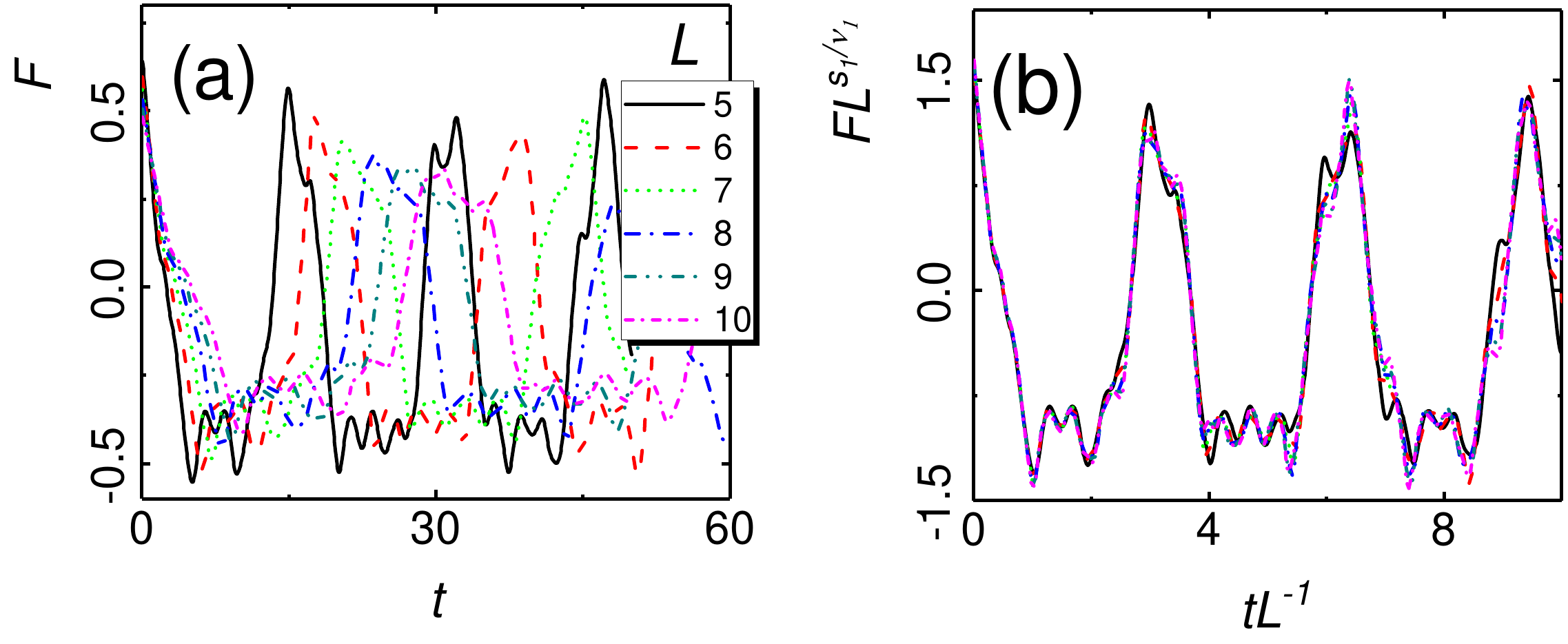}
  \caption{(a) The short-time dynamics of the OTOC for various $L$ with fixed $(\lambda-\lambda_c)L^{1/\nu_1}=0.02$ and $g_L L^{\beta_1\delta_1/\nu_1}=0.20444 i$. Note that the longitudinal field in model~(\ref{model}) has been chosen to be imaginary, which gives real longitudinal field as the usual quantum Ising model. (b) The rescaled curves collapse onto each other, confirming that Eq.~(\ref{isingu}) is also applicable for real parameters.}
\label{realotoc}
\end{figure}

Note that Eq.~(\ref{isingu}) should be applicable as long as the parameter is near the Ising critical point of model~(\ref{model}). We verify that Eq.~(\ref{isingu}) can also be used to describe the OTOC for the usual Hermitian quantum Ising model. In Fig.~\ref{realotoc}, we confirm this conclusion by calculating the model~(\ref{model}) with real external field.

\begin{figure}[tbp]
\includegraphics[angle=0,scale=0.4]{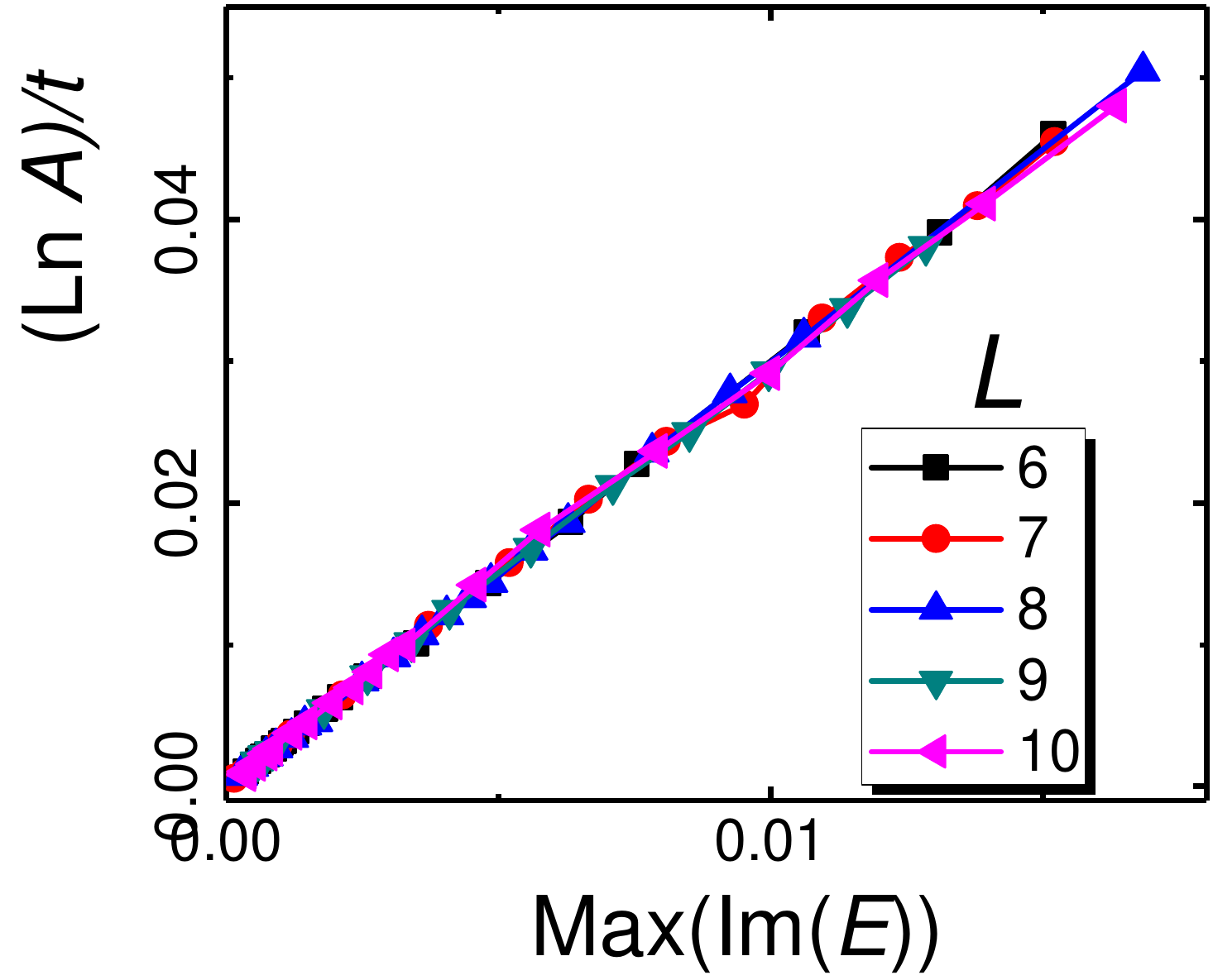}
  \caption{The curve of $({\rm Ln}A)/t$ versus $\textrm{Max}(\textrm{Im}(E))$ for various $L$. These curves depend scarcely on the lattice size. The slope is about $3$.}
\label{deps}
\end{figure}

\textit{Long-time dynamics controlled by the DEP}--- We now study the long-time dynamics of the OTOC. As shown in Fig.~\ref{general}, its amplitude increases exponentially. Here we point out the long-time dynamics is mainly determined by the states with maximum value of the imaginary-part of the energy spectra after the DEP. As shown in Fig.~\ref{deps}, the exponential fitting demonstrates that the OTOC satisfies $\mathcal{F}(t)\propto {\rm exp}[3\textrm{Max}(\textrm{Im}(E))t]$ in the long-time stage. To understand the coefficient in the exponent, we inspect Eq.~(\ref{OTOC}). The leftmost evolution operator ${\rm exp}(i\mathcal{H}t)$ acts on the ground state from the left side, and thus does not contributes to the divergent behavior, while each of three other evolution operators contributes one ${\rm exp}[\textrm{Max}(\textrm{Im}(E))t]$, since the imaginary-parts of the energy spectra always appear in pairs and only the positive part contributes to the long-time OTOC.

\textit{Discussion}--- Here we remark on the results. (a) The imaginary counterpart of the OTOC defined in Eq.~(\ref{OTOC}) also satisfies the conclusions obtained above. For example, one can find that the imaginary-part of Eq.~(\ref{twolev}) also obey the $(0+1)$D YLES scaling relation Eq.~(\ref{zeroscal}). (b) The OTOC can also be defined with a PT-symmtry breaking initial state. In this case, there is no oscillation stage. However, near the GEP, its short-time dynamics still satisfies the scaling equations~(\ref{zeroscal}) and (\ref{isingu}).

Recently, the non-Hermitian systems have been realized in experiments. For instance, the YLES has been detected in a open quantum system with a probe spin and an Ising bath~\cite{Wei,Peng}. Also, the critical dynamics of the PT-symmetric phase transition has been observed in a single-photon interferometric network~\cite{Xue1} and in a single spin system~\cite{Duj}. Furthermore, measurements of the OTOC have been implemented in various systems~\cite{dujf,rey}. Therefore, it is expected that the evolution of the OTOC can be measured in these experiments and the results obtained in this paper can also be detected therein.

\textit{Summary}--- We have studied the behavior of the OTOC in the non-Hermitian quantum Ising model. The role played by the DEP in the evolution of the OTOC has been identified. Near the GEP, the scaling behavior of the OTOC has been obtained. We have shown that the scaling behavior can be described by both the $(0+1)$D YLES and the $(1+1)$D Ising universality classes. The critical exponents of the OTOC in both universality classes have been determined. We have also pointed out that the OTOC scaling theory in the $(1+1)$D Ising universality class can be generalized into real-parameter cases. Moreover, in the long-time stage, we have found that the amplitude of the OTOC increases exponentially and the coefficient in its exponent is three times larger than the maximum value of the imaginary-part of the energy spectra. Although it is well-known that the GEP hosts the YLES, the property of the exceptional point in the excited states, to the best of our knowledge, has rarely been investigated. Our present work has taken the first step in this direction. Besides, it is also interesting to study the OTOC in the topological phase transitions of non-Hermitian systems~\cite{Harari}, and in the non-Hermitian disordered systems~\cite{Ueda3}. These are leaved for further works.

\textit{Acknowledgments}--- We would like to thank S.-K. Jian, S.-X. Zhang, Z. Yan, and Y. Huang for their helpful discussions. LJZ is supported by National Natural science Foundation of China (Grant No. 11704161) and the Natural Science Foundation of Jiangsu Province (Grant No. BK20170309).  SY is supported in part by China Postdoctoral Science Foundation (No. 2017M620035).


\end{CJK*}

\begin{widetext}

\newpage
\section{Supplemental Material}
\renewcommand{\theequation}{S\arabic{equation}}
\setcounter{equation}{0}
\renewcommand{\thefigure}{S\arabic{figure}}
\setcounter{figure}{0}
\renewcommand{\thetable}{S\arabic{table}}
\setcounter{table}{0}
In the supplemental material we study properties of the OTOC defined in a ferromagnetic state, $|\psi\rangle_\uparrow=\prod_i |\uparrow\rangle_i$.

First, we show the distribution of this wave function on the eigenstates in Fig.~\ref{diseig}. From Fig.~\ref{diseig}, one finds that the dominate parts are in the low-energy levels. Thus, the short-time dynamics of the OTOC of the $|\psi\rangle_F$ should be similar to the short-time dynamics of the OTOC defined in the ground state. Figure~\ref{zerodsup} shows that the near the GEP, the OTOC of the ferromagnetic state also satisfies Eq.~(\ref{zeroscal}). In addtion, Fig.~\ref{onedsup} shows that near the critical point, the OTOC of the ferromagnetic state satisfies Eq.~(\ref{isingu}).

\begin{figure}[bp]
\includegraphics[angle=0,scale=0.45]{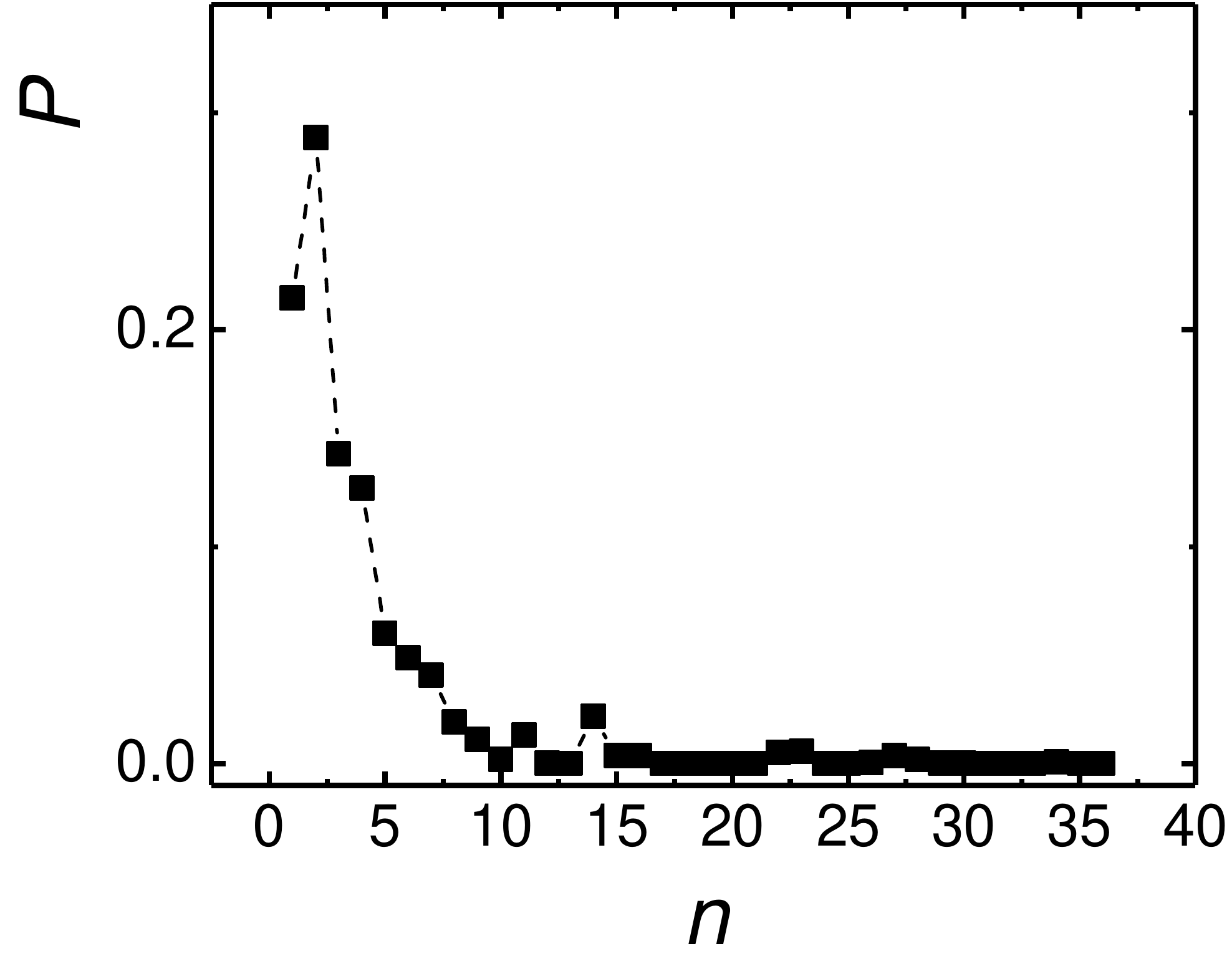}
  \caption{The distribution of $|\psi\rangle_\uparrow$ on the eigenstates for $L=8$, $\lambda=1.0025$, $h=0.00305$ and $k=0$. $P$ is defined as $P=|\langle E_n|\psi\rangle|^2$ and the index $n$ for the eigenstates is sorted in an ascending order according to the real-part of the eigenenergy.}
\label{diseig}
\end{figure}

\begin{figure}[bp]
\includegraphics[angle=0,scale=0.6]{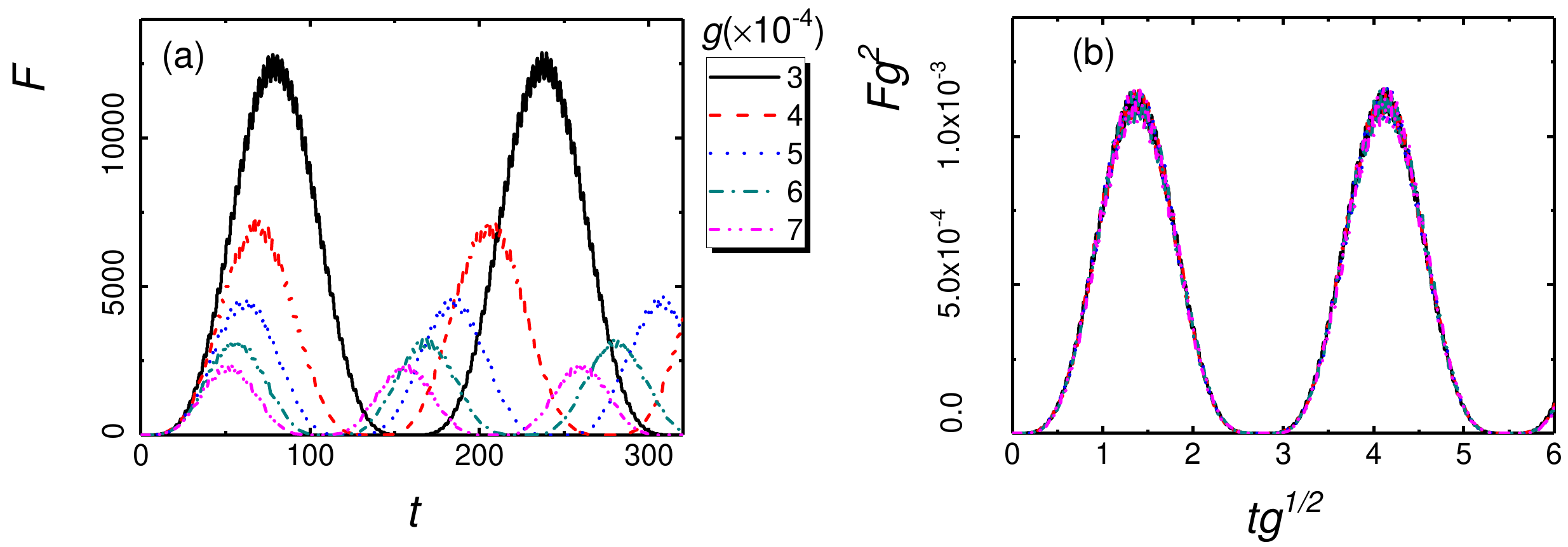}
  \caption{(a) The short-time dynamics of the OTOC in the ferromagnetic state for various $g$ with fixed $L=5$, and $\lambda=1.004$. The corrsponding $h_g$ is at $h_g=0.02785$. (b) The rescaled curves collapse onto each other, confirming that the OTOC in the ferromagnetic state still satisfies Eq.~(\ref{zeroscal}).}
\label{zerodsup}
\end{figure}

\begin{figure}[bp]
\includegraphics[angle=0,scale=0.6]{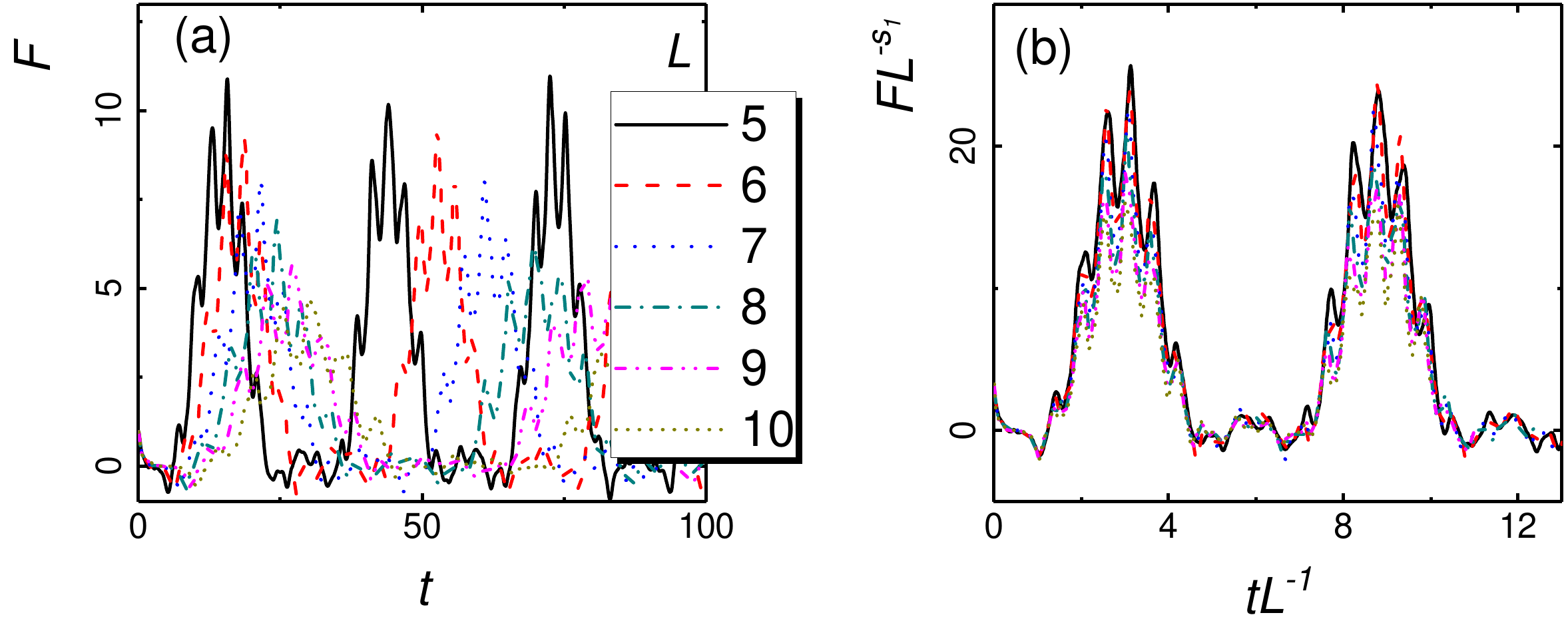}
  \caption{(a) The short-time dynamics of the OTOC in the ferromagnetic state for various $L$ with fixed $(\lambda-\lambda_c)L^{1/\nu_1}=0.02$ and $g_L L^{\beta_1\delta_1/\nu_1}=0.20444$. (b) The rescaled curves collapse onto each other, confirming that the OTOC in the ferromagnetic state also satisfies Eq.~(\ref{isingu}) with the same critical exponent $s_1$.}
\label{onedsup}
\end{figure}

Then, we show that the long-time dynamics of the OTOC in the ferromagnetic state is different from that in the ground state. Figure~\ref{longsup} shows although in the long-time scale, the OTOC still increases exponentially, the coefficient in the exponent is four times larger than the maximum imaginary-part of the energy spectra. The reason is that besides the middle three evolution operator, the leftmost one still contributes a ${\rm exp}({\rm Max}({\rm Im}(E)))$.

\begin{figure}[bp]
\includegraphics[angle=0,scale=0.4]{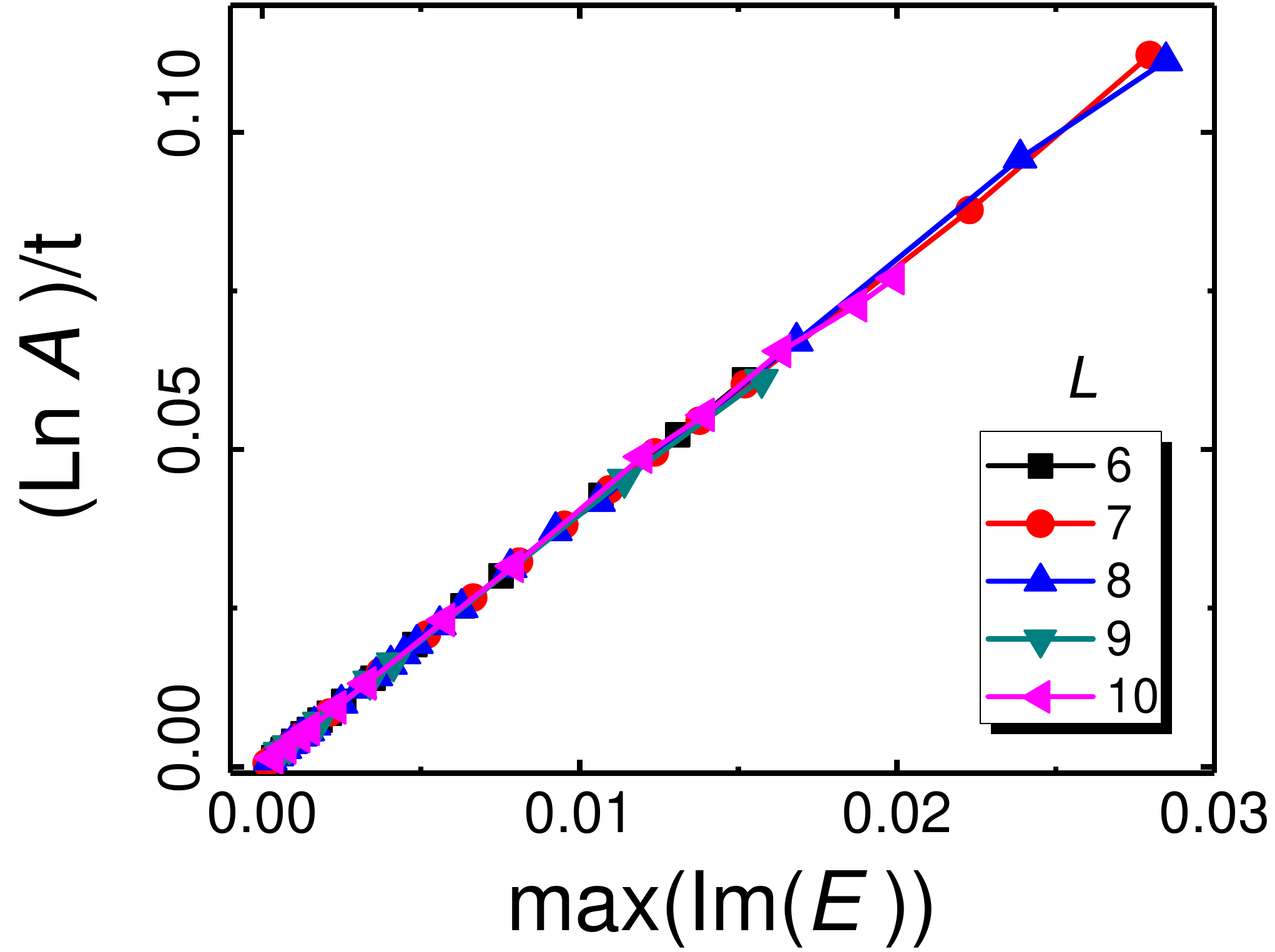}
  \caption{The curve of $({\rm Ln}A)/t$ versus $\textrm{Max}(\textrm{Im}(E))$ for the OTOC in the ferromagnetic state with various $L$. The slope is about $4$.}
\label{longsup}
\end{figure}

\end{widetext}


\begin{thebibliography}{99}
\bibitem{Dz} J. Dziarmaga, Adv. Phys. {\bf 59}, 1063 (2010).
\bibitem{Pol} A. Polkovnikov, K. Sengupta, A. Silva, and M. Vengalattore, Rev. Mod. Phys. {\bf 83}, 863 (2011).
\bibitem{Rig} L. D'Alessio, Y. Kafri, A. Polkovnikov, and M. Rigol, Adv. Phys. {\bf 65}, 239 (2016).
\bibitem{Abanin} D. A. Abanin, E. Altman, I. Bloch, and M. Serbyn, Rev. Mod. Phys. {\bf 91}, 021001 (2019).
\bibitem{Deu} J. M. Deutsch, Phys. Rev. A, {\bf 43}, 2046 (1991).
\bibitem{Sre} M. Srednicki, Phys. Rev. E, {\bf 50}, 888 (1994).
\bibitem{Rig1} M. Rigol, V. Dunjko, and M. Olshanii, Nature {\bf 452}, 854 (2008).
%
\bibitem{Basko}D. M. Basko, I. L. Aleiner, and B. L. Altshuler, Ann. Phys. (Amsterdam) {\bf 321}, 1126 (2006).
\bibitem{Oganesyan} V. Oganesyan and D. A. Huse, Phys. Rev. B {\bf 75}, 155111 (2007).
\bibitem{Schreiber} M. Schreiber, S. S. Hodgman, P. Bordia, H. P. L\"{u}schen, M. H. Fischer, R. Vosk, E. Altman, U. Schneider, and I. Bloch, Science {\bf 349}, 842 (2015).
\bibitem{Tur} C. J. Turner, A. A. Michailidis, D. A. Abanin, M. Serbyn and Z. Papi\'{c}, Nature Physics, {\bf 14}, 745 (2018).
\bibitem{Tur1} C. J. Turner, A. A. Michailidis, D. A. Abanin, M. Serbyn, and Z. Papi\'{c}, Phys. Rev. B {\bf 98}, 155134 (2018).
\bibitem{Khe} V. Khemani, C. R. Laumann, and A. Chandran, Phys. Rev. B {\bf 99}, 161101 (2019).
\bibitem{Lin} C.-J. Lin and A. Motrunich, Phys. Rev. B {\bf 122}, 173401 (2019).
\bibitem{Heyl} M. Heyl, A. Polkovnikov, and S. Kehrein, Phys. Rev. Lett. {\bf 110}, 135704 (2013).
\bibitem{Wei} B.-B. Wei and R.-B. Liu, Phys. Rev. Lett. {\bf 109}, 185701 (2012).
\bibitem{Mitra1} A. Mitra, Phys. Rev. Lett. {\bf 109}, 260601 (2012).
%
\bibitem{Berges2008} J. Berges, A. Rothkopf, and J. Schmidt, Phys. Rev. Lett. {\bf101}, 041603 (2008).
\bibitem{Gasenzer2011} B. Nowak, D. Sexty, and T. Gasenzer, Phys. Rev. B {\bf84}, 020506 (2011).
\bibitem{Erne2018} S. Erne, R. B\"{u}cker, T. Gasenzer, J. Berges, and J. Schmiedmayer, Nature {\bf563}, 225 (2018).
\bibitem{Oberthaler2018} M. Pr\"{u}fer, P. Kunkel, H. Strobel, S. Lannig, D. Linnemann, C. Schmied, J. Berges, T. Gasenzer, and M. K. Oberthaler, Nature {\bf563}, 217 (2018).
\bibitem{Eigen2018} C. Eigen, J. A. P. Glidden, R. Lopes, E. A. Cornell, R. P. Smith, and Z. Hadzibabic, Nature {\bf563}, 221 (2018).
%
\bibitem{Larkin} A. Larkin and Y. N. Ovchinnikov, Sov. Phys. JETP 28, 1200 (1969).
\bibitem{Shenker1} S. H. Shenker and D. Stanford, J. High Energy Phys. 03 (2014) 067.
\bibitem{Shenker2} S. H. Shenker and D. Stanford, J. High Energy Phys. 12 (2014) 046.
\bibitem{Maldacena} J. Maldacena, S. H. Shenker, and D. J. Stanford, J. High Energy Phys. 08 (2016) 106.
\bibitem{Herq} R.-Q. He and Z.-Y. Lu, Phys. Rev. B {\bf 95}, 054201 (2017).
%
\bibitem{Zhai} H. Shen, P. Zhang, R. Fan, and H. Zhai, Phys. Rev. B {\bf96}, 054503 (2017).
\bibitem{Heyl2019} M. Heyl, F. Pollmann, and Bal\'{a}zs D\'{o}ra, Phys. Rev. Lett. {\bf121}, 016801 (2018).
\bibitem{Duan} C. B. Da\v{g}, K. Sun and L.-M. Duan, arXiv: 1902.05041.
\bibitem{Sun} Z.-H. Sun, J.-Q. Cai, Q.-C. Tang, Y. Hu, and H. Fan, arXiv: 1811.11191.
\bibitem{Yang1952} C. Yang and T. Lee, Phys. Rev. {\bf87}, 404 (1952).
\bibitem{Lee1952} T. Lee and C. Yang, Phys. Rev. {\bf87}, 410 (1952).
\bibitem{Fisher}  M. Fisher, Phys. Rev. Lett. {\bf40}, 1610 (1978).
\bibitem{Peng} X. Peng, H. Zhou, B. B. Wei, J. Cui, J. Du, and R. B. Liu, Phys. Rev. Lett. {\bf114}, 010601 (2015).
\bibitem{Bender} C. M. Bender and S. Boettcher, Phys. Rev. Lett. {\bf80}, 5243 (1998).
\bibitem{Bender1} C. Bender, Rep. Prog. Phys. {\bf70}, 947 (2007).
\bibitem{Ueda1} Y. Ashida, S. Furukawa, M. Ueda, Nature Communications {\bf8}, 15791 (2017).
%
\bibitem{Harari}G. Harari, M. A. Bandres, Y. Lumer, M. C. Rechtsman, Y. D. Chong, M. Khajavikhan, D. N. Christodoulides, and M. Segev, Science {\bf359}, eaar4003 (2018).
\bibitem{Fuliang} H. Shen, B. Zhen, and L. Fu, Phys. Rev. Lett. {\bf 120}, 146402 (2018).
\bibitem{Fuliang2} H. Shen and L. Fu, Phys. Rev. Lett. {\bf 121}, 026403 (2018).
\bibitem{Wang} S. Yao and Z. Wang, Phys. Rev. Lett. {\bf121}, 086803 (2018).
\bibitem{Ueda2}Z. Gong, Y. Ashida, K. Kawabata, K. Takasan, S. Higashikawa, M. Ueda, Phys. Rev. X {\bf8}, 031079 (2018).
\bibitem{Yin1} S. Yin, G.-Y. Huang, C.-Y. Lo, and P. Chen, Phys. Rev. Lett. {\bf118}, 065701 (2017).
\bibitem{Yin2} L.-J. Zhai, H.-Y. Wang, and S. Yin, Phys. Rev. B {\bf97}, 134108 (2018).
\bibitem{Moessner} B. D\'{o}ra, M. Heyl, R. Moessner, Nature Communications, {\bf10}, 2254 (2019).
\bibitem{Ueda3} R. Hamazaki, K. Kawabata, M. Ueda, arXiv: 1811.11319.
\bibitem{Xue1} L. Xiao, K. Wang, X. Zhan, Z. Bian, K. Kawabata, M. Ueda, W. Yi, P. Xue, arXiv: 1812.01213.
\bibitem{Duj}Y. Wu, W. Liu, J. Geng, X. Song, X. Ye, C.-K. Duan, X. Rong, J. Du, arXiv: 1812.05226.
\bibitem{Sachdev} S. Sachdev, \textit{Quantum Phase Transitions}(Cambridge University Press, 2011).
\bibitem{supp} Supplemental material
\bibitem{weibobo} B.-B. Wei, G. Sun, and M.-J. Hwang, arXiv: 1906.00533.
\bibitem{dujf}J. Li, R. Fan, H. Wang, B. Ye, B. Zeng, H. Zhai, X. H. Peng, and J. F. Du, Phys. Rev. X {\bf 7}, 031011 (2017).
\bibitem{rey} M. G\"{a}rttner, J. G. Bohnet, A. Safavi-Naini, M. L. Wall, J. J. Bollinger, and A. M. Rey, Nat. Phys. {\bf 13}, 781 (2017).





\end{thebibliography}
\end{document}